\documentstyle[12pt,epsf]{article}
\newif\ifFigureInclude \FigureIncludefalse
\FigureIncludetrue
\ifFigureInclude\input{epsf}
\else{\typeout{(Figure won't be included)}}\fi

\newcommand{\AmS}{{\protect\the\textfont2
  A\kern-.1667em\lower.5ex\hbox{M}\kern-.125emS}}

\newcommand{\eq}[1]{Eq.~(\ref{eq:#1})}
\newcommand{\fig}[1]{Fig.~\ref{fig:#1}}

\catcode`@=11 
\def\lsim{\mathrel{\mathpalette\@versim<}}
\def\gsim{\mathrel{\mathpalette\@versim>}}
\def\@versim#1#2{\lower0.2ex\vbox{\baselineskip\z@skip
\lineskip\z@skip\lineskiplimit\z@\ialign{$\m@th#1
\hfil##\hfil$\crcr#2\crcr\sim\crcr}}}

\makeatletter
\def\setcaption#1{\def\@captype{#1}}
\makeatother
\newcommand{\figurehere}[4]{%
\begin{center}
\setcaption{figure}
\epsfxsize=#3 \epsfbox{#2}
\caption{#4}\
\label{fig:#1}
\end{center}
}
\newcommand{\beq}{\begin{equation}}
\newcommand{\eeq}{\end{equation}}
\def\bsub{\begin{mathletters}}
\def\esub{\end{mathletters}}

\newcommand{\beqa}{\begin{eqnarray}}
\newcommand{\eeqa}{\end{eqnarray}}

\newcommand {\av}{\mbox{{\scriptsize av}}}
\newcommand {\tot}{\mbox{{\scriptsize tot}}}
\newcommand {\lat}{\mbox{{\scriptsize lat}}}

\newcommand {\dd}{\mbox{d}}

\begin{document}

    \begin{normalsize}
     \begin{flushright}
                 UT-724\\
                 DPNU-95-29\\
                 hep-lat/9511023\\
                 Sep. 1995\\
{}~~\\
{}~~\\
     \end{flushright}
    \end{normalsize}
    \begin{Large}
       \vspace{1cm}
       \begin{center}
\renewcommand{\thefootnote}{\fnsymbol{footnote}}
         {\Large
Singular Vertices in the Strong Coupling Phase
of Four--Dimensional Simplicial Gravity}\footnote[2]{
based on the talk given at Lattice '95 in Melbourne and
``The new trend in the quantum field theory'' in Kyoto.}\\
\setcounter{footnote}{0}
\renewcommand{\thefootnote}{\arabic{footnote}}
       \end{center}
    \end{Large}

  \vspace{10mm}

\begin{center}
           Tomohiro H{\sc otta}$^{a)}$\footnote
           {
e-mail address : {\tt hotta@danjuro.phys.s.u-tokyo.ac.jp}},
           Taku I{\sc zubuchi}$^{a)}$\footnote
           {
talker of the two workshops,
e-mail address : {\tt izubuchi@danjuro.phys.s.u-tokyo.ac.jp} }
{\sc and} Jun N{\sc ishimura}$^{b)}$\footnote
           {e-mail address : {\tt nisimura@eken.phys.nagoya-u.ac.jp}}\\
      \vspace{1cm}
        $^{a)}$
        {\it Department of Physics, University of Tokyo ,} \\
        {\it Bunkyo-ku, Tokyo 113, Japan}\\
        $^{b)}$
        {\it  Department of Physics, Nagoya University,}\\
        {\it Chikusa-ku, Nagoya 464-01, Japan.}

\vspace{15mm}

\begin{abstract}
\noindent
\setlength{\baselineskip}{5mm}
We study four--dimensional simplicial gravity through numerical simulation
with special attention to
the existence of singular vertices, in the strong coupling phase,
that are shared by abnormally large
numbers of four--simplices.
The second order phase transition from the strong coupling phase
into the weak coupling phase could be understood as
the disappearance of the singular vertices.
We also change the topology of the universe from the sphere to the
torus.
\end{abstract}
\end{center}
\vfil\eject

\setcounter{footnote}{0}

\section{Introduction}
  One of the most exciting challenges in the theoretical physics is
to understand gravitational interaction in the context of the quantum
theory.
The problem we encounter when we try to formulate quantum gravity within
ordinary field theory in four dimensions is
that we cannot renormalize it perturbatively.
If we use lattice regularization, which enables a nonperturbative
study, general coordinate invariance is not
manifest and whether it is restored in the continuum limit is a crucial
problem.
One possibility of lattice regularization of quantum gravity is
dynamical triangulation, which is believed to restore
general coordinate invariance in the continuum limit.
It has been solved exactly in two dimensions
and its
continuum limit is shown to reproduce Liouville theory,
in which general coordinate invariance has been
treated carefully.

Although four--dimensional dynamical triangulation
seems to be difficult to solve analytically, there is no potential barrier
in studying it through numerical simulation.
Employing the Einstein--Hilbert action as the lattice action and sweeping
the gravitational constant,
it has been discovered that the system undergoes a second order
phase transition \cite{4DQG,Brug93a}, which suggests the possibility
of taking a continuum limit.
One of the main purpose of this paper is to try to clarifying the physical
meaning of this phase transition.

When we regularize four--dimensional quantum gravity
with dynamical triangulation the integration over the metric
is replaced with the random summation over
all possible four--dimensional simplicial manifolds.

Although we can modify the lattice action expecting universality,
it is natural  to start with the Euclidean Einstein--Hilbert action
\begin{equation}
  S = \int \dd^4  x \sqrt{g} \left( \Lambda - \frac{1}{G} R \right)
\label{eq:EHaction}
\end{equation}
as a first trial,
where $\Lambda$ is the cosmological constant and $G$ is the
gravitational constant.
Let us denote the number of $i$-simplices in a simplicial manifold by
$N_i$.
One can easily find that, for a simplicial manifold,
\beqa
\int \dd^4 x \sqrt{g}  &=& c N_4 \\
\int \dd^4 x \sqrt{g} R  &=& 2 \pi N_2 - 10 \alpha N_4 ,
\eeqa
where $c$ is the volume of each four--simplex and
$\alpha$ is the angle between two faces of a four--simplex,
which is equal to $\arccos \left(\frac{1}{4}\right)$.
Therefore the Einstein--Hilbert action (\ref{eq:EHaction}) can be
expressed in terms of lattice variables as
\begin{equation}
  S_{\lat} = \kappa_4 N_4 - \kappa_2 N_2 ,
  \label{eq:lataction}
\end{equation}
where $\kappa_4$ and $\kappa_2$ are related to $\Lambda$ and $G$ through,
\begin{equation}
  \kappa_4=  c \Lambda + \frac{10 \alpha}{G},~~~~~~~~~~~~~
  \kappa_2=  \frac{2 \pi}{G}.
\label{eq:kappa2G}
\end{equation}

\section{The Vertex Order Concentration}
We consider an ensemble with a fixed $N_4$ and
with spherical topology.
There are well established methods for generating
such an ensemble through numerical simulations,
and the technical details of our simulation shall be given elsewhere
\cite{HIN}.
Our code is written for arbitrary dimension following Ref. \cite{Catt94}.



Let us turn to the results of our simulation.
We first look at the second order phase transition, which can be seen
through thermodynamic quantities such as the average curvature per
unit volume :
%
\beqa
R_{\av} &=& \frac{R_{\tot}}{N_4} ~~~
(R_{\tot}= \int \dd^4  x \sqrt{g} R)  \\
       &=&  2 \pi \frac{N_2}{N_4} - 10 \alpha .
\eeqa
\label{eq:Rave}
\fig{fig1} shows our results for $\langle R_{\av} \rangle$
at various $\kappa_2$'s.

\figurehere{fig1}{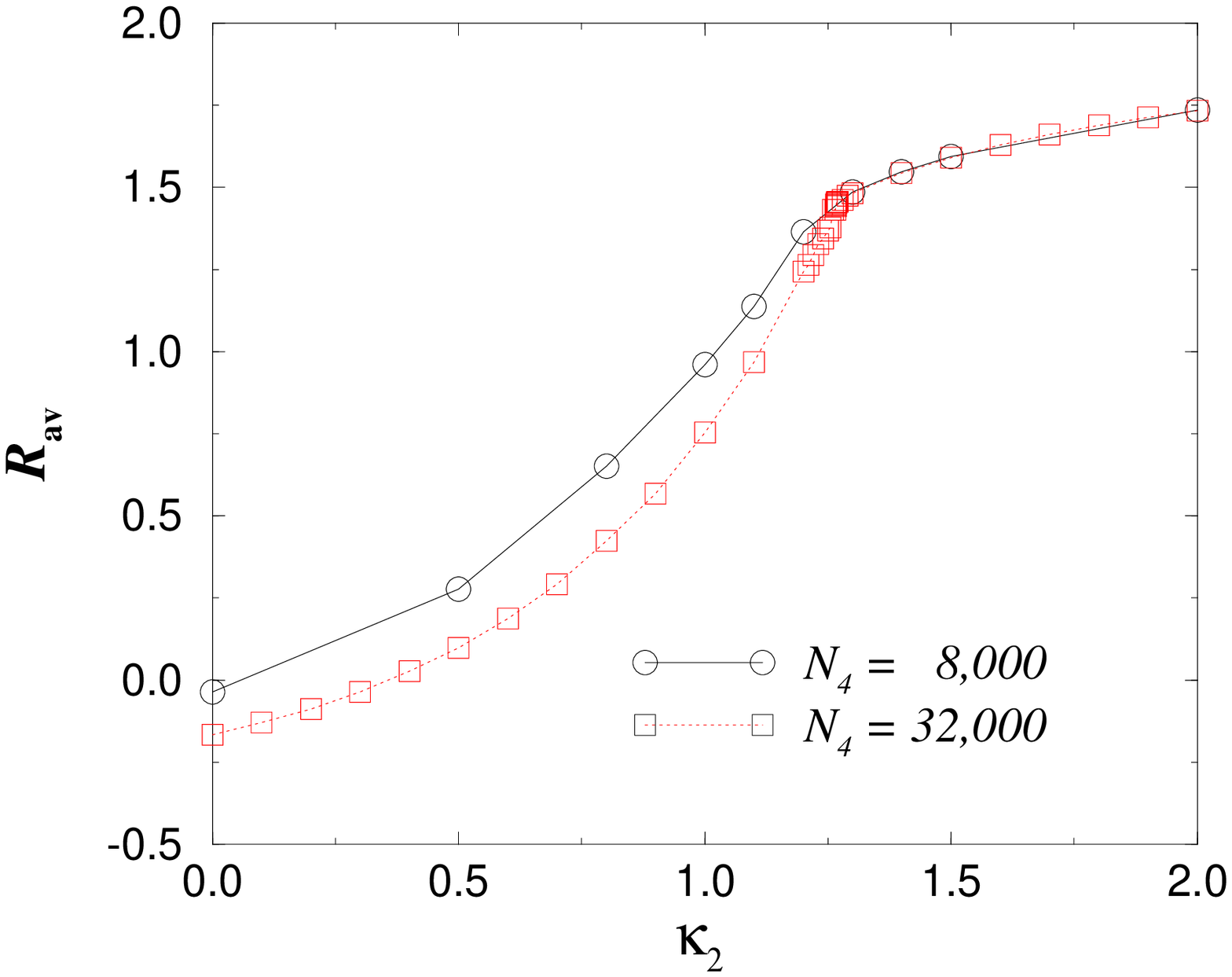}{12cm}{ The average curvature per volume is
plotted against \protect$\kappa_2$ for \protect$N_4=8000, 32000$.}

In contrast to the three--dimensional case \cite{3DQG},
no hysteresis has been observed.
Also, one sees that the size dependence of the data changes abruptly
at $\kappa_2 = 1.2 \sim 1.3$. On the right there is little size
dependence, whereas on the left, the curve goes lower and lower as
we increase the system size.
The derivative of the average curvature gives the susceptibility
\begin{equation}
\chi_R = 2 \pi  \frac{\partial\langle R_{\av}\rangle}{\partial \kappa_2}
     = \frac{\langle R_{\tot}^2 \rangle -
       \langle R_{\tot}\rangle^2}{N_4} ,
\end{equation}
which represents the fluctuation of the total curvature.
As is expected from \fig{fig1},
the susceptibility has a peak
around $\kappa_2 = 1.2 \sim 1.3$, which grows higher as the
system size is increased.
%
%
This implies that the correlation length of the local curvature diverges
at the critical point \cite{twopoint}, where we may hope to take a
continuum limit.
Since $\kappa_2$ corresponds to the inverse of the gravitational constant,
as is seen from (\ref{eq:kappa2G}), we call the large $\kappa_2$ phase as
the {\it weak coupling phase}\/
and the small $\kappa_2$ phase as the {\it strong coupling phase}\/ .

Although this phase transition has been observed by
many authors\cite{4DQG,Brug93a}, the physical
origin of this transition might be not understood clearly.

To clarify it, we measure the vertex order distribution as follows.
Let the vertex order $o(v)$  be the number
of four-simplices sharing the vertex $v$.
Then the vertex order distribution $\rho$ can be defined as
\beq
\rho(n) \equiv \frac{1}{N_0} \langle \sum _{v} \delta_{o(v),n} \rangle.
\label{eq:rhoDef}
\eeq
This quantity is measured every 100 sweeps and averaged
over 100 configurations.
In order to reduce the fluctuations of the distribution,
we smear the data over bins of size 10.

We first note that the average vertex order per one vertex,
$\overline{o(v)}$, can be given as
\beq
\overline{o(v)}
\equiv \frac{1}{N_0} \sum_v o(v) = \frac{5 N_4}{N_0} =
{10 N_4 \over N_2- 2 N_4 + 4}.\label{eq:AvOv}
\eeq
In the second equality, we used the relation
$\sum_v o(v) = 5 N_4,$
which comes from the fact that each four-simplex has five vertices.

{}From \eq{lataction} and \eq{AvOv},
when we move from the strong coupling phase ($\kappa_2 \sim 0.0$) to
the weak coupling
phase ($\kappa_2\sim 2.0$) with fixed system size,
$N_2$ increases and thus $\overline{o(v)}$ goes to a smaller value.

In \fig{fig3} we show the vertex
order distribution $\rho(n)$ for $\kappa_2$ = 0.0, 1.267
(near the critical point) and $2.0$ with $N_4=32,000$.
For $\kappa_2=0.0$,
one finds that there is an isolated peak of {\it very large vertex order}\/;
as large as one third of the total four-simplices.

As $\kappa_2$ is increased from 0.0 the position of the peak shifts to
left in accordance with the decrease of the $\overline{o(v)}$
and around the critical point ($\kappa_2 \sim 1.2$)
the peak is absorbed into the continuum part of the distribution.
For $\kappa_2=2.0$, on the other hand, the distribution damps
quite rapidly for large vertex order and we have confirmed that
it remains almost unchanged when we increase the system size.
\figurehere{fig3}{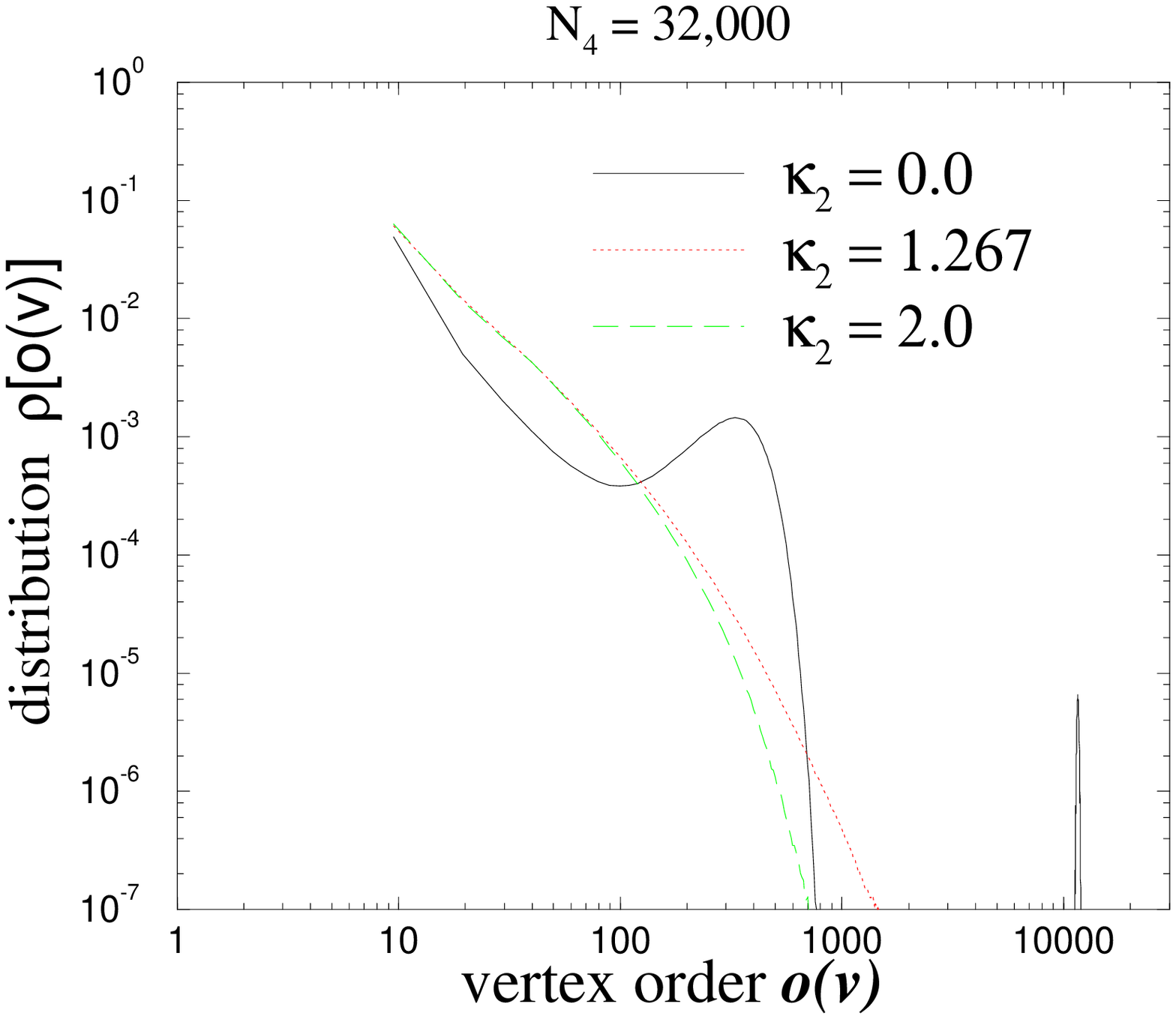}{12cm}{
    The vertex order distribution for $\kappa_2=0, 1.267, 2.0$
    for $N_4=32000$.}
In \fig{lam0_N4} we show the size dependence of the vertex order
distribution for $\kappa_2=0.0$.
One finds that the very large vertex order grows linearly
as one increases $N_4$,
and thus this concentration might be relevant even
in the thermodynamic limit.

We call this peculiar phenomenon as the
{\it vertex order concentration (VOC)}\/.
We have also confirmed that the peak consists of two vertices%
\footnote{ For D-dimensional dynamical triangulation($3 < D < 7$),
it seems that the peak consists of $D-2$ vertices.},
and we found no link order concentration; there is no singular link shared
by conspicuously large numbers of four--simplices.
\figurehere{lam0_N4}{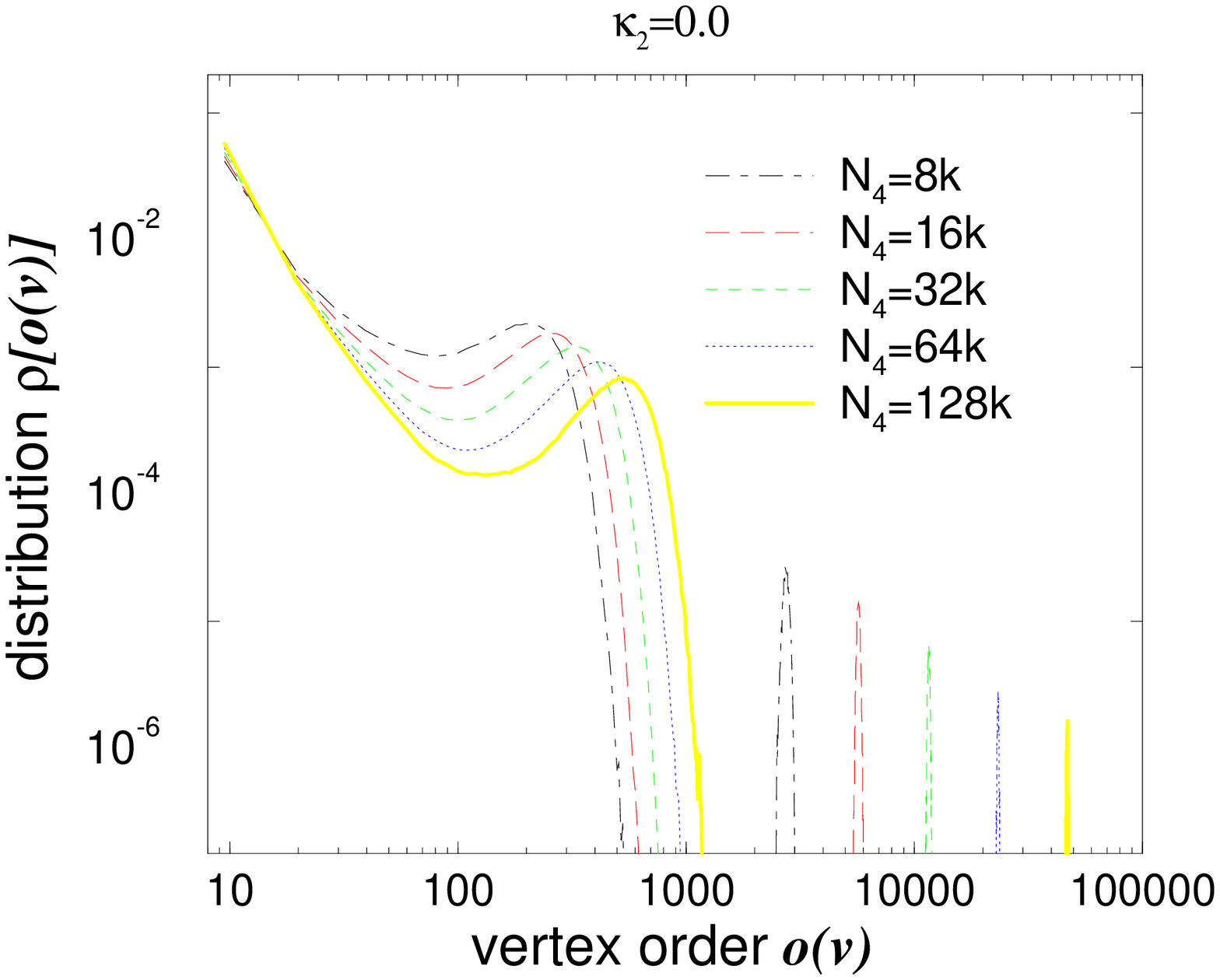}{12cm}{
    The vertex order distribution for $\kappa_2=0$
    varying the system size $N_4$ from 8k to 128k.}
\section{The thermalization check}
One may suspect that our Monte Carlo simulation is trapped
by a local minimum (meta-stable state)
and sweeps over only a limited part of the whole
configuration space.
To check that this suspicion is not the case,
we prepare three types of configurations
for initial configurations from which we start simulations.

The first one consists of six four-simplices
which are the surface of a five-simplex;
we call this the {\it hot start}\/ configuration.
The second one is the  {\it cold start}\/, for which we prepare
approximately flat configuration, using the surface of the
five-dimensional rectangular complex (box).
The system size of the cold start configuration can be
adjusted near the target number $N_4^0$
and it has no VOC\cite{HIN}.
The third configuration, {\it four-VOC  configuration}\/, has
four singular vertices and is made in the following way.
We prepare two configurations of system size $N_4^0/2$
by performing sufficiently many sweeps in  strong coupling phase.
Each of the thermalized configurations has two singular vertices.
Then we identify one four-simplex of one configuration with
one four-simplex of the other, so that
the resulting manifold is the four-sphere of system size $N_4^0-1$
with four singular vertices.
So this configuration has twice as many singular vertices
as the configuration obtained in strong coupling region.

After performing  more than 10,000 Monte Carlo sweeps \footnote{we
define one sweep as $N_4^0$ times accepted updates.} we found no
dependence on the three types of initial configurations
for any of our observables.

For example, we show the history of the vertex order below.
Let us label the vertices of the configuration as $v_i$, so that
$o(v_i) \geq o(v_{i+1})$ is satisfied for any $i=1,2 \cdots N_0$.
We show the $o(v_1)$ (the largest vertex order) as a function of
the number of Monte Carlo sweep from each initial configuration
together with $o(v_2)$ and $o(v_3)$ for $\kappa_2=0.0$
in \fig{vohist}.
We only show the results of the simulation for the cold and hot starts
for the legibility.
For each types of initial configuration
(thick curves or  thin curves),
one can see that there are two large vertex orders
 $o(v_1)$ and $o(v_2)$
around $25,000$ and a large {\it gap}\/ between $o(v_3)$ and
the above two.
Below the curve of $o(v_3)$, the curves of $o(v_i), i\geq 3$
run closely without significant gaps.
One can easily see that the two curves of each $o(v_i), i=1,2,3$ are almost
identical after 10,000 sweeps.
\figurehere{vohist}{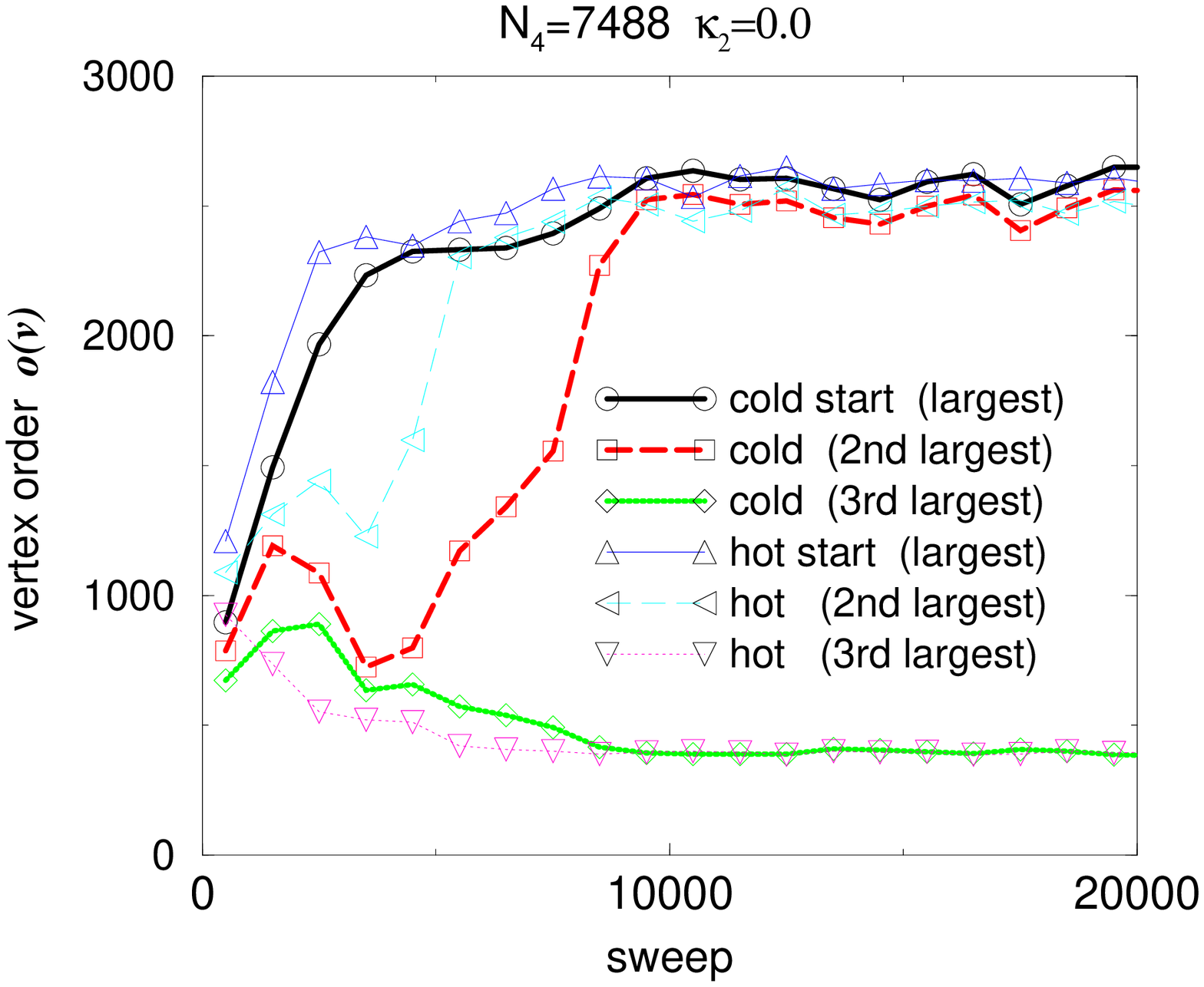}{12cm}{
    The history of the vertex order for $\kappa_2=0$
    for the hot and cold initial configurations.}
Considering that the vertex order distributions of the three types
of configurations are quite different from each other,
the above fact  strongly suggests that our result,
the existence of two
singular vertices in strong coupling region, does not come
from the insufficiency of thermalization but reflects true properties
of the path-integral measure of the dynamical triangulation.

\section{The torus topology}
  We test a possibility of the phenomenon of the VOC being related
to the constraint of the topology of the manifold,{\it i.e}\/
the manifold must be $S^4$.
We change the manifold from four-dimensional sphere ($S^4$) to
the torus ($(S^1)^4$).
The method of making the $(S^1)^4$ triangulation is as follows.
{
\renewcommand{\theenumi}{\Alph{enumi})}
\renewcommand{\labelenumi}{\theenumi}
\begin{enumerate}
\item Prepare the four-dimensional rectangular complex (four-box)
and identify each pair of parallel boundary.
Draw one line $l_d$ between two vertices which is diagonal
to each other.
\label{enu:i1}
\item Divide the four-boxes into $4!$ four-simplices.
    The dividing edges on boundaries of the four-boxes must
    be projections of the diagonal line of $l_d$ for
    consistent construction of the torus triangulation.
\end{enumerate}
}
For simplicity \fig{TorusTriang} describes the three-dimensional case.

\figurehere{TorusTriang}{torusFig.eps}{12cm}{
The construction of the torus triangulation.}

The resulting triangulation has $(S^1)^4$ topology, from which
we start the Monte Carlo simulation.

\fig{TorusRes} shows the comparison of the vertex order
distribution $\rho(n)$ between the sphere and the torus case
for $N_4=128,000$. The $\rho(n)$ of the torus of the
size $N_4\lsim 100,000$ is sensitive to whether the manifold is
the torus or the sphere, while  for the larger manifold
$N_4\sim 128,000$ the distribution is almost identical
between the two topologies (\fig{TorusRes}).

\figurehere{TorusRes}{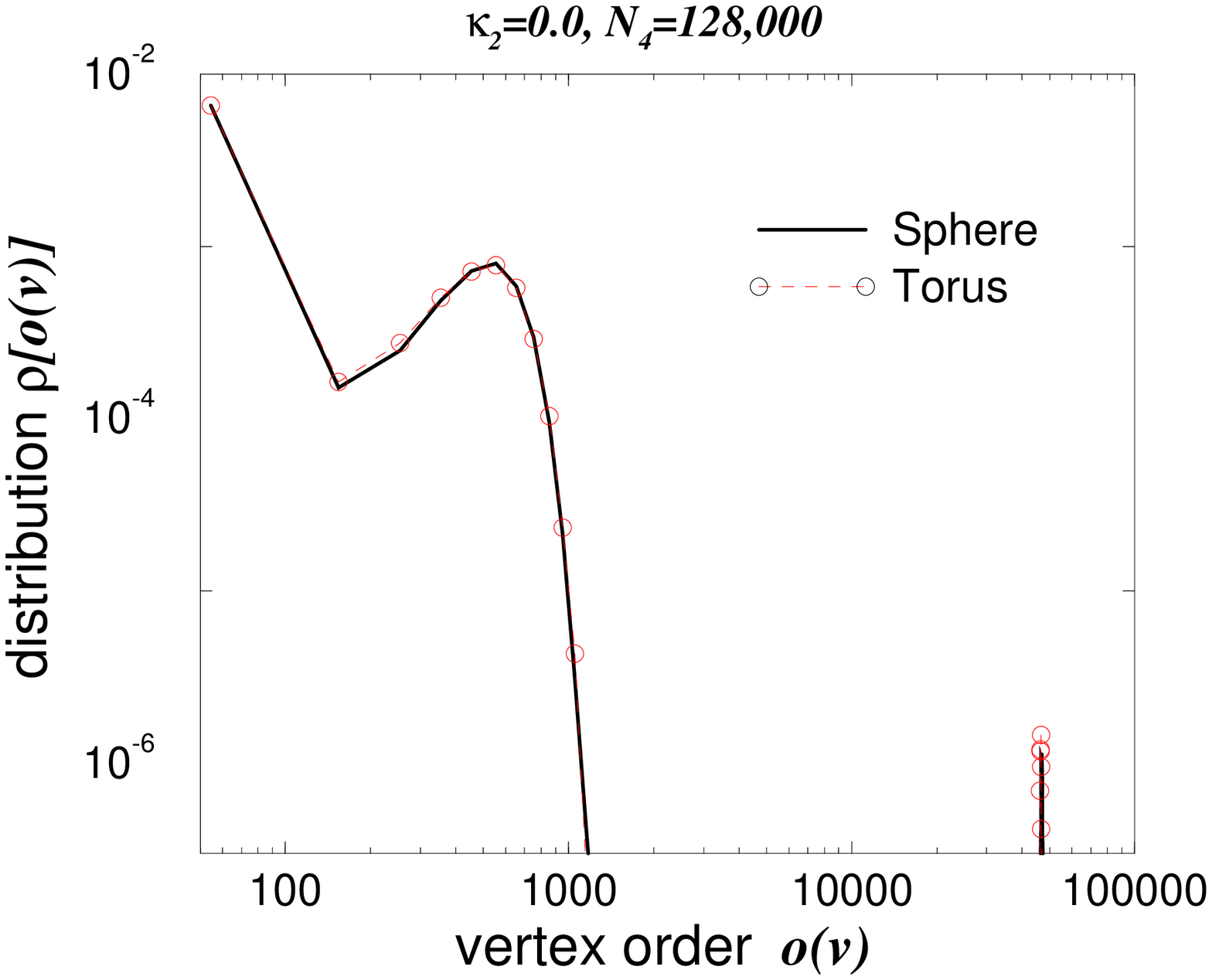}{12cm}{
The vertex order distribution of the sphere and the torus.
$N_4=128,000, \kappa_2=0.0$.}

The number of the singular vertices in torus manifold is
one for $N_4\lsim 100,000$ and two for $N_4\sim 128,000$.
{}From these results we might say that the finite size effect of
torus is severer than the sphere case.
And the vertex order distribution $\rho(n)$ seems to be almost
insensitive to the topology difference, but of course
more high statistic simulations of more variation of topology
({\it e.g.}\/ $(S^2)^2$, $D^4$ {\it etc.}\/) are necessary for
the definite conclusions.

\section{Discussion}
  To summarize, we found two singular vertices in the
strong coupling region, which disappear in weak coupling region.
Although the thermalization is checked,
the existence of this  VOC is still very strange.
The VOC in $\kappa_2=0$ case, for which we
observe purely the measure of path integral
without any weight from the action, implies that the number of
such VOC configuration is much larger than the
number of the smooth (non VOC)  configuration in
dynamical triangulation.
Such a singular behavior might be an obstacle to
the continuum limit.
Considering universality in quantum gravity,
as well as in ordinary field theories,
we think that a sound second order phase transition without VOC,
where we can take a sensible continuum limit, should be searched
by modifying the lattice action\cite{Amb92a,Brug93a,HIN}
to suppress the VOC.

\par

{\large\bf  Acknowledgements.}

\vspace*{0.5cm}

  Numerical calculations for the present work have been performed on
HP-700 series at KEK and Univ. of Tokyo.
  We would like to thank Prof. H.~Kawai, Prof. T.~Yukawa and Dr. N.~Tsuda
for stimulative discussions.
  This work is supported in part by the Grants-in-Aid of the
Ministry of Education.

\end{document}